\documentclass[
onecolumn
]{aastex631}
\usepackage{graphics} % for pdf, bitmapped graphics files
\usepackage{epsfig} % for postscript graphics files
\usepackage{mathptmx} % assumes new font selection scheme installed
\usepackage{times} % assumes new font selection scheme installed
\usepackage{amsmath} % assumes amsmath package installed
\usepackage{amssymb}  % assumes amsmath package installed
\usepackage{xcolor}
\usepackage{subfigure}
\usepackage{bm}
\usepackage{hyperref}
\graphicspath{{./}}

\submitjournal{ApJS}

\shorttitle{Re-binned vs. Spectral Binned Energetic Particle Intensities}
\shortauthors{Cuesta et al.}

\begin{document}

% \title{\bf Clarifying the Integration and Averaging of Particle Intensity in Logarithmic Space}
% \title{Re-binned versus Spectral Binned Particle Intensities}
\title{Comparing Methods for Calculating Solar Energetic Particle Intensities: Re-binning versus Spectral Binning}

\author[0000-0002-7341-2992]{M. E. Cuesta}
\affiliation{Department of Astrophysical Sciences,
Princeton University, Princeton, NJ 08544, USA}
\correspondingauthor{Manuel Enrique Cuesta}
\email{mecuesta@princeton.edu}

\author[0000-0003-0412-1064]{L. Y. Khoo}
\affiliation{Department of Astrophysical Sciences, Princeton University, Princeton, NJ 08544, USA}

\author[0000-0002-7655-6019]{G. Livadiotis}
\affiliation{Department of Astrophysical Sciences, Princeton University, Princeton, NJ 08544, USA}

\author[0000-0002-3093-458X]{M. M. Shen}
\affiliation{Department of Astrophysical Sciences, Princeton University, Princeton, NJ 08544, USA}

\author[0000-0003-2685-9801]{J. R. Szalay}
\affiliation{Department of Astrophysical Sciences, Princeton University, Princeton, NJ 08544, USA}

\author[0000-0001-6160-1158]{D. J. McComas}
\affiliation{Department of Astrophysical Sciences, Princeton University, Princeton, NJ 08544, USA}

\author[0000-0002-8111-1444]{J. S. Rankin}
\affiliation{Department of Astrophysical Sciences, Princeton University, Princeton, NJ 08544, USA}

\author[0000-0002-6962-0959]{R. Bandyopadhyay}
\affiliation{Department of Astrophysical Sciences, Princeton University, Princeton, NJ 08544, USA}

\author[0000-0001-7952-8032]{H. A. Farooki}
\affiliation{Department of Astrophysical Sciences, Princeton University, Princeton, NJ 08544, USA}

\author[0000-0001-6286-5809]{J. T. Niehof}
\affiliation{Space Science Center, University of New Hampshire, Durham, NH 03824, USA}

\author[0000-0002-0978-8127]{C. M. S. Cohen}
\affiliation{Space Research Lab, California Institute of Technology, Pasadena, CA 91125, USA}

\author[0000-0002-0156-2414]{R. A. Leske}
\affiliation{Space Research Lab, California Institute of Technology, Pasadena, CA 91125, USA}

\author[0000-0002-9246-996X]{Z. Xu}
\affiliation{Space Research Lab, California Institute of Technology, Pasadena, CA 91125, USA}

% \author[0000-0003-0581-1278]{G. D. Muro}
% \affiliation{Space Research Lab, California Institute of Technology, Pasadena, CA 91125, USA}

% \author[0000-0002-5674-4936]{M. E. Hill}
% \affiliation{Applied Physics Laboratory, Laurel, MD 20723, USA}

% \author[0000-0003-1960-2119]{D. G. Mitchell}
% \affiliation{Applied Physics Laboratory, Laurel, MD 20723, USA}

\author[0000-0003-2134-3937]{E. R. Christian}
\affiliation{NASA Goddard Space Flight Center, Greenbelt, MD 20771, USA}

\author[0000-0002-7318-6008]{M. I. Desai}
\affiliation{Southwest Research Institute,
San Antonio, TX 78238, USA}
\affiliation{University of Texas,
San Antonio, TX 78249, USA}

\author[0000-0001-9323-1200]{M. A. Dayeh}
\affiliation{Southwest Research Institute,
San Antonio, TX 78238, USA}
\affiliation{University of Texas,
San Antonio, TX 78249, USA}

%% Note that the \and command from previous versions of AASTeX is now
%% depreciated in this version as it is no longer necessary. AASTeX 
%% automatically takes care of all commas and "and"s between authors names.

%% Mark off the abstract in the ``abstract'' environment. 
\begin{abstract}
Solar energetic particle (SEP) events have been observed for decades in the interplanetary medium by spacecraft measuring the intensity of energetic ions and electrons. These intensities provide valuable information about particle acceleration, the effects of bulk plasma dynamics on particle transport, and the anisotropy of particle distributions. Since measured intensities are typically reported in narrow energy bins, it is common to re-bin intensities over a wider energy range to improve counting statistics. We investigate two methods for calculating intensities across multiple energy bins: a) \textit{re-binned intensity} (\(\overline{j}_{\rm linlin}\)), which is calculated by integrating the intensity over energy space and corresponds to the intensity at an effective energy that depends on the time-varying spectral index, and b) \textit{spectral binned intensity} (\(\overline{j}_{\rm loglog}\)), calculated by integrating the log-intensity in log-energy space, yielding the intensity at the log-centered energy that is independent of the spectral index and remains constant over time. We compare these methods using Parker Solar Probe (PSP) IS\(\odot\)IS measurements of energetic protons, and we prescribe criteria for selecting the appropriate method for different scenarios. Our results show that the re-binned intensity is consistently larger (up to a factor of 5) than the spectral binned intensity for two SEP events observed by PSP, although the time series of the two methods are strongly correlated. Overall, both measures are important for SEP spectral analysis, and the selection of the appropriate measure depends on whether a physical (spectral binned intensity) or a statistical (re-binned intensity) representation is needed for a given analysis.
    \newline
\end{abstract}

%% Keywords should appear after the \end{abstract} command. 
%% See the online documentation for the full list of available subject
%% keywords and the rules for their use.

%% From the front matter, we move on to the body of the paper.
%% Sections are demarcated by \section and \subsection, respectively.
%% Observe the use of the LaTeX \label
%% command after the \subsection to give a symbolic KEY to the
%% subsection for cross-referencing in a \ref command.
%% You can use LaTeX's \ref and \label commands to keep track of
%% cross-references to sections, equations, tables, and figures.
%% That way, if you change the order of any elements, LaTeX will
%% automatically renumber them.
%%
%% We recommend that authors also use the natbib \citep
%% and \citet commands to identify citations.  The citations are
%% tied to the reference list via symbolic KEYs. The KEY corresponds
%% to the KEY in the \bibitem in the reference list below. 

%%%%%%%%%%%%%%%%%%%%%%%%%%%%%%%%%%%%%%%%%%
%%%%%%%%%%%%%%%%%%%%%%%%%%%%%%%%%%%%%%%%%%
\section{Introduction} \label{sec:intro}
%%%%%%%%%%%%%%%%%%%%%%%%%%%%%%%%%%%%%%%%

Solar energetic particles (SEPs) routinely permeate the solar wind and can be accelerated by various mechanisms.
For example, SEPs can be accelerated by interplanetary shocks via turbulent diffusion acceleration processes \citep{Fisk1971JGR_CosRayIntensity_IPshock,AxfordEA1977ICRC_CosRayAccel_Shock,BlandfordOstriker1978ApJL_ParticleAccel_Shock,Bell1978MNRAS_ShockAccel_CosRay,GoslingEA1979AIPC_IonAccel_EarthBowShock,Lee1983JGR_IonAccel_IPshock,Lee1997GMS_CMEshock_transport}.
These processes are typically seeded by the suprathermal particle population with energies ranging from \(\sim10~{\rm keV}\) to \(\sim1~{\rm MeV}\) per nucleon \citep{DayehEA2009ApJ_SuprathermalSeedPopulation,MasonGloeckler2012SSRv_suprathermals}.
Since the early 1960s, SEP observations in the solar wind have been widely studied using particle detectors onboard spacecraft.
These particle detectors measure the count rates of incident particles (${\rm N}/\Delta t$), where \({\rm N}\) is the number of detected particles over a time interval (\(\Delta t\)). 
For each instrument-dependent geometric factor and efficiency (\(G_\textrm{eff}\)), the intensity of an arbitrary energy bin can be calculated via \(j = {\rm N}/G_\textrm{eff}\Delta tq\Delta E\) with typical units of (cm$^{2}$ sr s eV)$^{-1}$, where $q$ is the total charge of the particle and $\Delta E$ is the width of the energy bin in linear space.
%Using the geometric factor of each specific detector and the incident particle's energy, quantities such as differential `\(Q\)' flux (\(QdN/d\Omega dtdE\)).For instance, the differential energy flux is \(EdN/d\Omega dtdE\) with units of \({\rm eV/(cm^2\cdot sr \cdot s \cdot eV)}\) for \(Q=E\) can be derived.Here we deal with differential number flux, which is \(dN/d\Omega dtdE\) with units of \({\rm (cm^2\cdot sr \cdot s \cdot eV)^{-1}}\) for \(Q=1\).Throughout this paper, we use the common term for this -- intensity.

The intensity is an important quantity in space research pertaining to energetic particles, since it provides information regarding their temporal evolution, spatial variation, and even energization. When count rates are very low, computing an ``average" intensity, with respect to time, energy, or between different apertures for example, is commonly done to improve counting statistics.
Here, we only consider the case of averaging intensity with respect to energy.
Calculating an energy average in the linear space of both the intensity and energy is the widely used method for characterizing the intensity over a range of energy \citep[e.g., ][]{RankinEA2021ApJ_PSP_FirstACRs,DresingEA2023AandA_WidespreadSEPevent_17April2021,BucikEA2023AandA_SolO_FirstGradualEPevent_He3rich,CuestaEA2024ApJ_CMEShockCorrelation,KouloumvakosEA2024AandA_MultiSC_EPs_28Oct2021,ZhuangEA2024ApJ_SEPAccelReleaseCoronalShock,SchwadronEA2024ApJ_PSP_EPs_CMEflank}.
This procedure is often referred to as re-binning the intensity since the originally constructed energy bins are cast into a larger energy bin 
% (larger \(\Delta E/E\))
, effectively summing the number of particles detected within the wider bin.
This \textit{re-binned intensity} should be assigned to the effective energy, but the log-centered energy is used with the understanding that there is some error, which has been examined for the construction of individual energy bins \citep[][]{KronbergDaly2013GIMDS_WideEnergySpectralAnalysis}, but not yet for combined energy bins, to our knowledge.

The effective energy associated with the re-binned intensity varies over time, and only in special circumstances does the effective energy equal the log-centered energy, which does not vary over time.
Although the ``bow-tie" method can be used to calculate an effective energy that remains nearly constant in time when combined with the geometry factor \citep{SelesnickBlake2000JGR_RadiationBeltRelativisticElectrons_Bowtie,ClaudepierreEA2021SSRv_MagneticElectronIonSpectrometer_Bowtie}, the convenience of the log-centered energy is generally preferred since power law spectra can be fitted linearly in the logarithmic space of energy and intensity.
Therefore, to answer questions related to comparisons over time, the intensity along the spectrum at the log-centered energy, or the \textit{spectral binned intensity}, is a more appropriate measure since the intensities correspond to the same log-centered energy.
The spectral binned intensity requires an energy average in the logarithmic space of both the intensity and energy.
As a result, we investigate the effects of averaging the intensity over linear vs. logarithmic space and we provide a list of various realistic scenarios to help determine which method is more appropriate to use.

In this paper, we investigate the re-binned intensity in comparison to the spectral binned intensity.
We present analytic expressions for both measures and explore their ratio, which depends on three parameters: the spectral index, the energy range, and the original energy bin width in logarithmic space.
Regarding observations, we examine a series of SEP events observed by the Parker Solar Probe \citep[PSP; ][]{Fox2016SSRv_PSP} and compare the re-binned and spectral binned intensities for protons measured by the Integrated Science Investigation of the Sun \citep[IS\(\odot\)IS; ][]{McComasEA2016SSRv_ISOIS} instrument suite.
The paper is organized as follows. In Section \ref{sec:methods}, we explain the methods (Section \ref{subsec:formulation}) and derive analytic expressions for the re-binned and spectral binned intensities (Section \ref{subsec:analytic}) used to reach the modeled results in Section \ref{sec:modeled} and results from spacecraft observations in Section \ref{sec:results}.
We summarize and discuss these results in Section \ref{sec:discussion}, including the underlying representation of the two different measures of particle intensity investigated here.

%For practical demonstration, we compute the re-binned and spectral binned intensities for energetic protons observed by the Integrated Science Investigation of the Sun \citep[IS\(\odot\)IS; ][]{McComasEA2016SSRv_ISOIS} instrument suite onboard the Parker Solar Probe \citep[PSP; ][]{Fox2016SSRv_PSP} spacecraft during an SEP event.We find that the re-binned intensity and the spectral binned intensity have their own strengths and weaknesses.For instance, the re-binned intensity can overestimate the spectral binned intensity by nearly a factor of 5 (assuming both intensities are used to characterize the log-centered energy); however, their time series are strongly correlated (information regarding the qualitative evolution of the intensity as a function of time is retained).In cases where one wishes to compare intensities corresponding to the same energy between different times and/or spacecraft, then the spectral binning method is more appropriate.This is analogous to comparing the intensity of the spectrum at a particular energy.Additionally, it is more convenient to use the spectral binning method to fit intensities over time or energy (in logarithmic space) without introducing further error unnecessarily.On the other hand, the re-binning method can be used as a measure of the total number of particles within an energy range, but it does not represent the  spectrum of any fixed energy.

\section{Methodology} \label{sec:methods}

\subsection{Basic Formulation} \label{subsec:formulation}

We focus on the intensity \(j(E)\) at an arbitrary time throughout this study, which is constructed for an individual energy bin as \(j = {\rm N}/G_\textrm{eff}\Delta tq\Delta E\).
Although the intensity normalized by the energy bin width in linear space (\(\Delta E\)), the spacing of the energy bins themselves may be constructed in other spaces that are nonlinear, such as square-root or logarithmic.
Here, we assume all energy bins have equal widths in logarithmic space. 
% We assume the intensity is constructed in the linear space of energy (\(\Delta E\)).
% This means that for an originally constructed energy bin, \(j = {\rm N}/G_\textrm{eff}\Delta tq\Delta E\).
% However, the grid spacing of energy bins may be constructed in other spaces that are nonlinear, such as square-root or logarithmic.
%For the purposes of this paper, we only focus on energy bins whose widths are equal in logarithmic space.
% The reason for this stems from the common observation that the number of counts at higher energy is much smaller compared to that at lower energy for falling spectra.
% Even for flat or rising spectra, particle counts when combined with the geometry factor and the incident particle's energy may still be lower at higher energy than lower energy.
Energetic particle instruments typically collect count rates in logarithmically spaced energy bins, and such spacing is convenient in both gathering more counting statistics at higher energy and in analyzing the spectra in logarithmic space, where fits to the spectra are often linear. The overall goal of the methods discussed below is to combine multiple bins of energetic particle measurements into a single bin.

\begin{figure}[ht]
    \centering
    \includegraphics[width=.5\linewidth]{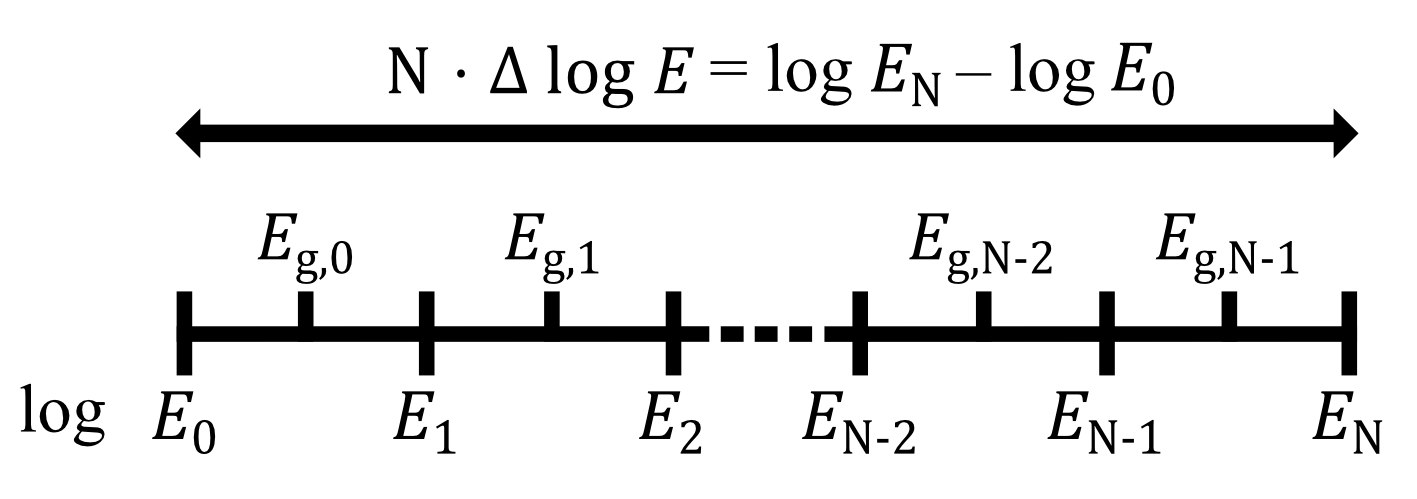}
    \caption{An illustration explaining the grouping of discrete, non-overlapping energy bins in logarithmic space for a number of \(N\) energy bins.}
    \label{fig:energy_grid}
\end{figure}

First, we revisit the procedure used to combine energy bins to calculate an average intensity of the energy range represented by a single wider energy bin.
Figure \ref{fig:energy_grid} illustrates an energy grid of discrete, non-overlapping bins logarithmic space\footnote{In cases where the energy bins are not adjacent and may overlap, a different treatment may be required for the analytic representation of the re-binned and spectral binned intensities.} that we wish to re-bin (i.e, by merging \(N\) number of bins).
The bin edges of the original bins run from \(E_0\) to \(E_N\), where the maximum edge of one bin is the minimum edge of the subsequent bin (adjacent bins), with \(E_{{\rm g},i}\) denoting the geometric, or log-centered, energy of the \(i\)-th energy bin.

%\newpage

%The energy average in the `y'-space of an arbitrary independent variable (\(F\)) constructed in the `x'-space of energy can be written as:
The average intensity over a specific energy range is given by,
\begin{equation} \label{eq:average}
    \overline{j}_{\rm xy} ={\rm \hat{Y}}^{-1} \left[\langle {\rm \hat{Y}}(j)\rangle_{{\rm \hat{X}}(E)}\right]={\rm \hat{Y}}^{-1} \left[ \frac{\int {\rm \hat{Y}}(j)d({\rm \hat{X}}(E))}{\int d({\rm \hat{X}}(E))} \right]\text{,}
\end{equation}
where \({\rm \hat{Y}}\) is the function that transforms the space for averaging \(j\) separately from the function \({\rm \hat{X}}\) that describes the space of energy, \({\rm \hat{Y}}^{-1}\) is the inverse function of \({\rm \hat{Y}}\) that is used to transform back to the original space after averaging, and \({\rm xy}\) are the corresponding labels of the average describing the functions \({\rm \hat{X}}\)  and \({\rm \hat{Y}}\).
For instance, the average of the intensity over \textit{linear} space with energies described in \textit{linear} space gives \({\rm xy}={\rm ``linlin"}\).
On the other hand, the average of the intensity over \textit{logarithmic} space with energies described in \textit{logarithmic} space gives \({\rm xy}={\rm ``loglog"}\).
These are the only two cases that we consider throughout this paper, but other combinations are possible.
The formalism for the inclusion of \({\rm \hat{Y}}\) and its inverse function has been proven by \cite{Aczel1948BAMS_AverageInverseFunctions} if \({\rm \hat{Y}}\) is a continuous and strictly increasing function (the logarithm function satisfies these conditions) and as been demonstrated elsewhere in the literature \citep[e.g., ][]{Livadiotis2007_FittingGeneralMethods,LivadiotisMcComas2012ApJ_KappaThermoProcesses}.
% As a result, setting \({\rm xy}={\rm linlin}\), where \({\rm \hat{X}}(E)=E\) and \({\rm \hat{Y}}(F)=F\), for a linear description of both \(F\) and \(E\).
% On the other hand, to describe both \(F\) and \(E\) in logarithmic space, then set \({\rm xy}={\rm loglog}\) where \({\rm \hat{X}}(E)=\log E\) and \({\rm \hat{Y}}(F)= \log F\).
% Here, we only consider cases where the average is labeled by \({\rm x}={\rm y}\in\{{\rm log,lin}\}\), or \({\rm \hat{X}} = {\rm \hat{Y}}\).

To determine the total counts within an energy range spanning multiple energy bins (statistically re-bin), we set \({\rm \hat{Y}}(j)=j\), \({\rm \hat{X}}(E)=E\), and \({\rm xy}={\rm ``linlin"}\) such that:

\begin{equation} \label{eq:j_linlin}
    \overline{j}_{\rm linlin} = \frac{\sum\limits^{N-1}_i j(E_{{\rm g},i}) \Delta E_i}{\sum\limits^{N-1}_i \Delta E_i} \cong \frac{\int jdE}{\int dE}
\end{equation}

where \(N\) is the number of discrete energy bins to be combined, \(E_{{\rm g},i}\) is the log-centered energy of the \(i\)-th energy bin with width \(\Delta E_i\) for \(i\) starting from zero.
Notice that we have assumed the intensities of the original energy bins are described at their corresponding log-centered energy.
In general, this is not true, since they are integrated counts divided by the width of the constructed energy bin \(\Delta E\).
However, in the case that the energy bin width divided by the minimum bin edge is much less than one or sufficiently small, then the error associated with our assumption in Eq. \eqref{eq:j_linlin} is small \citep{KronbergDaly2013GIMDS_WideEnergySpectralAnalysis}.
Alternatively, the spectral binned intensity is defined as the value along the spectrum at the log-centered energy (\(E_{\rm g}\)), or geometric mean energy.
This requires an arithmetic average in the logarithmic space of both the intensity and energy.
We refer to Eq. \eqref{eq:average} and set \({\rm \hat{Y}}(j) = \log j\), \({\rm \hat{X}}(E) = \log E\), and \({\rm xy}={\rm ``loglog"}\) such that:

\begin{equation} \label{eq:j_loglog}
    \overline{j}_{\rm loglog} = 10^{\sum\limits^{N-1}_i \log j(E_{{\rm g},i}) \Delta \log E_i / \sum\limits^{N-1}_i \Delta \log E_i} \cong 10^{\int \log j d(\log E)/\int d(\log E)}\text{.}
\end{equation}

The values of \(\overline{j}_{\rm linlin}\) and \(\overline{j}_{\rm loglog}\) represent the values of the observed spectrum at energies \(E_{\rm eff}\) and \(E_{\rm g}\), or \(j(E_{\rm eff})=\overline{j}_{\rm linlin}\) and \(j(E_{\rm g})=\overline{j}_{\rm loglog}\), respectively, where \(E_{\rm eff}\) is the effective energy.
Eqs. \eqref{eq:j_linlin} and \eqref{eq:j_loglog} represent the values for \(\overline{j}_{\rm linlin}\) and \(\overline{j}_{\rm loglog}\) calculated from observations.
However, we must first examine what parameters drive the differences between the re-binned and spectral binned intensities to fully understand their difference for observed SEP spectra.
Therefore, in Section \ref{subsec:analytic}, we derive analytic expressions for the values of \(\overline{j}_{\rm linlin}\), \(\overline{j}_{\rm loglog}\), \(E_{\rm eff}\), and \(E_{\rm g}\) assuming a single power law distribution, which are used as models for comparing \(\overline{j}_{\rm linlin}\) with \(\overline{j}_{\rm loglog}\) for simulated particle intensity spectra with various energy bin specifications in Section \ref{sec:modeled}.

\subsection{The Case of a Single Power Law Distribution} \label{subsec:analytic}

Now we examine Eqs. \eqref{eq:j_linlin} and \eqref{eq:j_loglog} assuming the observed intensity spectra can be described by a single power law distribution, such that \(j(E)=A\cdot E^{-\gamma}\) for some scaling coefficient \(A\) and spectral index (\(\gamma\)).
To find the value \(E_{\rm eff}\), we examine \(\overline{j}_{\rm linlin}\). Substituting \(j(E)\) into Eq. \eqref{eq:j_linlin}:

\begin{equation}\label{eq:linlin_sub1}
    \overline{j}_{\rm linlin} = \frac{\sum\limits^{N-1}_i (A\cdot E_{{\rm g},i})^{-\gamma} \Delta E_i}{\sum\limits{N-1}_i \Delta E_i}\text{.}
\end{equation}
For energy bins, described by Figure \ref{fig:energy_grid}, that are spaced equally in logarithmic space (\(\Delta \log E_{{\rm g},i}=\Delta \log E={\rm const.}\)), the \(i\)-th log-centered energy is given by:

\begin{equation} \label{eq:log_center_energy}
    \log E_{{\rm g},i} = \log E_{{\rm g},0} + i\cdot \Delta \log E \text{; or } E_{{\rm g},i} = E_{{\rm g},0}\cdot 10^{i\cdot \Delta \log E}\text{,}
\end{equation}
where \(E_0\) is the starting energy of the re-binned range.
Similarly, the width \(\Delta E_i\) can be expressed in terms of the logarithmic energy spacing as:

\begin{equation} \label{eq:log_energy_width}
    \Delta E_i = E_{i+1} - E_i = 10^{\log E_{{\rm g},i} + \frac{1}{2}\Delta \log E} - 10^{\log E_{{\rm g},i} - \frac{1}{2}\Delta \log E} = E_{{\rm g},0} \left( 10^{\frac{1}{2}\Delta \log E} - 10^{- \frac{1}{2}\Delta \log E} \right) 10^{i\cdot \Delta \log E}\text{.}
\end{equation}
Substituting Eqs. \eqref{eq:log_center_energy} and \eqref{eq:log_energy_width} into \eqref{eq:linlin_sub1} gives:

\begin{align}
    \overline{j}_{\rm linlin} &= A\cdot E_{{\rm g},i}^{-\gamma} \left( \frac{\sum\limits^{N-1}_i 10^{(1-\gamma)\cdot i \cdot \Delta \log E}}{\sum\limits^{N-1}_i 10^{i \cdot \Delta \log E}} \right) \\
    &= A\cdot E_{{\rm g},i}^{-\gamma} \left( \frac{10^{\Delta \log E} -1}{10^{N\cdot \Delta \log E} -1}\right) \left( \frac{10^{N(1-\gamma)\Delta \log E}-1}{10^{(1-\gamma)\Delta \log E} -1} \right) \\
    &= A\cdot E_{{\rm g},i}^{-\gamma} \left( \frac{\left(\frac{E_N}{E_0}\right)^{1-\gamma}-1}{\left(\frac{E_N}{E_0}\right)-1} \right) \left( \frac{10^{\Delta \log E}-1}{10^{(1-\gamma)\Delta \log E}-1} \right) \text{,} \label{eq:j_linlin_analytic}
\end{align}
or (for \(j(E_{\rm eff}=A\cdot E_{\rm eff}^{-\gamma}=\overline{j}_{\rm linlin}\)):

\begin{equation} \label{eq:eff_energy}
    E_{\rm eff} = E_{{\rm g},0} \left[ \left( \frac{\left(\frac{E_N}{E_0}\right)^{1-\gamma}-1}{\left(\frac{E_N}{E_0}\right)-1} \right) \left( \frac{10^{\Delta \log E}-1}{10^{(1-\gamma)\Delta \log E}-1} \right) \right]^{-1/\gamma}
\end{equation}
which is dependent on the spectral index over the re-binned energy range.
Since the spectral index can vary over time within any arbitrary energy range, the energy corresponding to \(\overline{j}_{\rm linlin}\) can also vary over time.
Consequently, comparing \(\overline{j}_{\rm linlin}\) at different times, especially over large time scales, may not be meaningful in some situations. 
The re-binned intensity is a useful measure of the number of particles observed within the specified energy range, as it effectively returns to count space, sums the counts over the wider energy range and is then normalized by the wider energy bin width.
Therefore, \(\overline{j}_{\rm linlin}\) can be used to determine whether the number of particles per geometry factor per time observed within the energy range has increased or decreased.
However, \(\overline{j}_{\rm linlin}\) is not reliable as a metric for the intensity of the spectrum characteristic of a fixed energy over an arbitrarily large time range, because the effective energy varies over time (spectral index varies over time).
Consequently, if the spectral index varies quickly, then a meaningful time scale of comparison for \(\overline{j}_{\rm linlin}\) shortens accordingly.

Now we examine the spectral binned intensity defined as the value along the spectrum at the log-centered energy.
Using the relation from Eq. \eqref{eq:log_center_energy} and representing \(j(E)\) as a single power law from Eq. \eqref{eq:j_loglog}:

\begin{align}
    \overline{j}_{\rm loglog} &= 10^{\sum\limits^{N-1}_i (\log A - \gamma \log E_{{\rm g},i})\delta \log E_i / \sum\limits^{N-1}_i \Delta \log E_i} \\
    &= A\cdot E_{{\rm g},0}^{-\gamma} \cdot 10^{-\frac{\gamma \cdot (N-1)}{2} \Delta \log E} \\
    &= A\cdot E_{{\rm g},0}^{-\gamma} \left( \frac{E_N}{E_0} \right)^{-\frac{\gamma}{2}}\cdot 10^{\frac{\gamma}{2}\Delta \log E}\text{i} \label{eq:j_loglog_analytic}
\end{align}
or (for \(j(E_{\rm g}=A\cdot E_{\rm g}^{-\gamma}=\overline{j}_{\rm loglog}\)):

\begin{equation}\label{eq:geo_energy}
    E_{\rm g} = E_{{\rm g},0} \cdot \left( \frac{E_N}{E_0} \right)^{\frac{1}{2}} \cdot 10^{-\frac{1}{2}\Delta \log E} = \sqrt{E_0E_N} \text{,}
\end{equation}
such that \(N\cdot \Delta \log E = \log E_N - \log E_0\) and \(E_{{\rm g},0} = E_0 10^{\frac{1}{2}\Delta \log E}\).
Note, \(E=E_{\rm g}\) is exactly the log-centered energy of the wider energy bin.
Also, it is completely independent of the spectral index and therefore will not vary over time unlike \(E_{\rm eff}\).
Therefore, \(\overline{j}_{\rm loglog}\) has the desired property of a spectral binned intensity, such that it is equivalent to the intensity along the spectrum at the fixed log-centered energy.

\begin{figure}[ht]
    \centering
    \includegraphics[width=.6\linewidth]{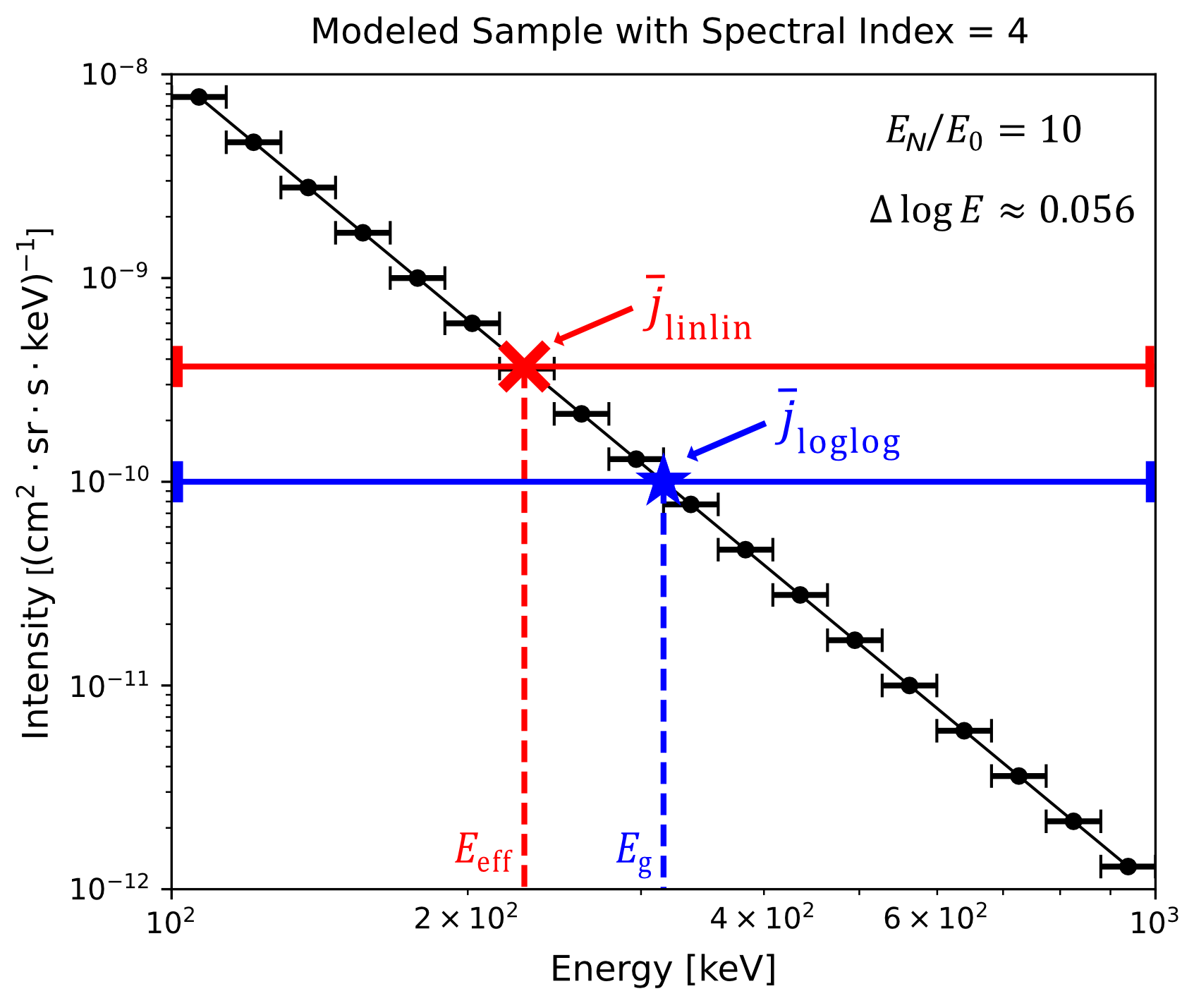}
    \caption{A simulated intensity-energy spectrum \(E^{-\gamma}\) with spectral index \(\gamma=4\). The red-\(\times\) and blue-\(\star\) mark the values of \(\overline{j}_{\rm linlin}\) and \(\overline{j}_{\rm loglog}\), respectively, calculated over 100 -- 1,000~keV.
    % in the left panel and over two different energy ranges in the right panel (marked by the thin dashed vertical black line). 
    Error bars for \(\overline{j}_{\rm linlin}\) and \(\overline{j}_{\rm loglog}\) mark the new energy bin from the combination of the original energy bins (black circles with their energy bin widths equal in logarithmic space). 
    % The horizontal red/blue dashed lines in the left panel represent the expected intensity characteristic of the spectrum at 
    The vertical red/blue dashed lines mark the effective/log-centered energy of the range used in the evaluation of \(\overline{j}_{\rm linlin}\) and \(\overline{j}_{\rm loglog}\), respectively. The original energy bins have a logarithmic spacing \(\Delta \log E \approx 0.056\) spanning a decade of energy.}
    \label{fig:motivation}
\end{figure}

To visualize the difference between \(\overline{j}_{\rm linlin}\) and \(\overline{j}_{\rm loglog}\), we provide a simulated intensity-energy spectrum constructed with energy bins of equal width in logarithmic space, as shown in Figure \eqref{fig:motivation}.
The spectrum is given a spectral index of 4 (e.g., \(\gamma = 4\) for \(E^{-\gamma}\)), a common value for the spectral index found to describe SEP spectra observed in the solar wind \citep[e.g., ][]{FiskGloeckler2006ApJL_CommonSpectrumSuprathermal,SchwadronEA2020ApJS_SeepPop_Accel_PSP,LeskeEA2020ApJS_SEP_April2019_PSP,CohenEA2021AandA_2020Nov29SEPevent_PSP,LivadiotisEA2024ApJ_KappaTechnique_SEPthermo,CuestaEA2024APJ_Thermodynamics_of_SEPs}.
The red-\(\times\) and blue-\(\star\) mark the values of \(\overline{j}_{\rm linlin}\) and \(\overline{j}_{\rm loglog}\) corresponding to Eqs. \eqref{eq:j_linlin} and \eqref{eq:j_loglog}, respectively, calculated over the entire energy range from 100 -- 1,000~keV.
The blue dashed line illustrate where \(\overline{j}_{\rm loglog}\) falls on the spectrum at the log-centered energy, reflecting the property of arithmetic averages in logarithmic space of variables that behave linearly in logarithmic space.
On the other hand, \(\overline{j}_{\rm linlin}\) describes the spectrum at the effective energy (dashed red line) that would vary as a function of the spectral index spectral index.
% On the right panel, we split the spectrum into a third of the decade on the left and two thirds of a decade on the right, consisting of a different number of combined energy bins.
The horizontal error bars belonging to \(\overline{j}_{\rm linlin}\) and \(\overline{j}_{\rm loglog}\) mark the resulting energy bin of the averaged energy range.
% Note that if \(\overline{j}_{\rm linlin}\) is used to represent the combined energy ranges at \(E_{\rm g}\), then the resulting spectral index over the new energy bins will decrease, whereas the spectral index remains unaffected if \(\overline{j}_{\rm loglog}\) is used.

\newpage

One complication that arises with Eq. \eqref{eq:j_loglog} is the existence of zero count measurements, a common occurrence in particle data reflecting real information (not missing data).
Since the uncertainty of measured zeros is the only physical tie to its surrounding points in the spectrum, we first replace any measured zero with its upper limit uncertainty (to avoid taking the logarithm of zero) before applying Eq. \eqref{eq:j_loglog} to the data.
Upper limit uncertainties are calculated according to \cite{Gehrels1986ApJ_PoissonStatistics} (see their Eq. (9) using a 84.13\% confidence level).
% \textbf{
We investigated the impact of choosing the upper limit uncertainty as the replaced value as opposed to a small fraction of the upper limit uncertainty, yielding insignificant impact on higher count times.
% }
This is expected since in the limit of high counting probability, the likelihood of acquiring a zero-count measure will be little to none.
% \textbf{
However, the impact on very low count times was strongly dependent on the choice of this fraction.
% }
Consequently, \(\overline{j}_{\rm loglog}\) is expected to be reliably unbiased in the high counting regime, whereas the bias of \(\overline{j}_{\rm loglog}\) increases with decreasing counting statistics (and vice versa).
For now, we compute \(\overline{j}_{\rm linlin}\) and \(\overline{j}_{\rm loglog}\) on the same modified spectra. 
% \textbf{
The treatment of zero intensity measurements is still an active area of improvement for future research.
% }

\section{Comparisons with modeled spectra} \label{sec:modeled}

The goal of this paper is to make the distinction in the utilization between the re-binned intensity (\(\overline{j}_{\rm linlin}\)) and the spectral binned intensity (\(\overline{j}_{\rm loglog}\)) and optimize the usage of each, depending on the scientific question being addressed.
Although it is conventional to describe the spectrum at the log-centered energy using \(\overline{j}_{\rm linlin}\), we have shown that its effective energy is dependent on the spectral index, unlike \(\overline{j}_{\rm loglog}\).
The ratio \(\overline{j}_{\rm loglog}/\overline{j}_{\rm linlin}\) will be used to determine when \(\overline{j}_{\rm loglog}\) should be applied in favor of \(\overline{j}_{\rm linlin}\), or when \(\overline{j}_{\rm linlin}\) strongly deviates from \(\overline{j}_{\rm loglog}\).
By using the expressions derived in the previous section, we can take the ratio of Eq. \eqref{eq:j_loglog_analytic} to Eq. \eqref{eq:j_linlin_analytic} such that:

\begin{equation} \label{eq:ratio}
    \frac{\overline{j}_{\rm loglog}}{\overline{j}_{\rm linlin}} = \left( \frac{10^{(1-\gamma)\Delta \log E}-1}{10^{\Delta \log E}-1} \right) \cdot \left( \frac{\left(\frac{E_N}{E_0}\right)-1}{\left(\frac{E_N}{E_0}\right)^{1-\gamma}-1} \right)\cdot \left(\frac{E_N}{E_0}\right)^{-\frac{\gamma}{2}} \cdot 10^{\frac{\gamma}{2}\Delta \log E}\text{,}
\end{equation}
which is a function of three parameters: the spectral index \(\gamma\), the range of the new energy bin \(\frac{E_N}{E_0}\), and the width of the original energy bins \(\Delta \log E\) (assumed to be constant for logarithmically spaced bins but can also be reworked for other energy bin spacings if desired).
Note that in the case that \(N=1\), then only one energy bin is considered, and the ratio is equal to 1.
This reflects the assumption that the intensities of the original bins are expressed at the log-centered energy, even though using \(\overline{j}_{\rm linlin}\) at his energy introduces little error for sufficiently small energy bin widths \citep{KronbergDaly2013GIMDS_WideEnergySpectralAnalysis}.

\begin{figure}[ht]
    \centering
    \includegraphics[width=0.8\linewidth]{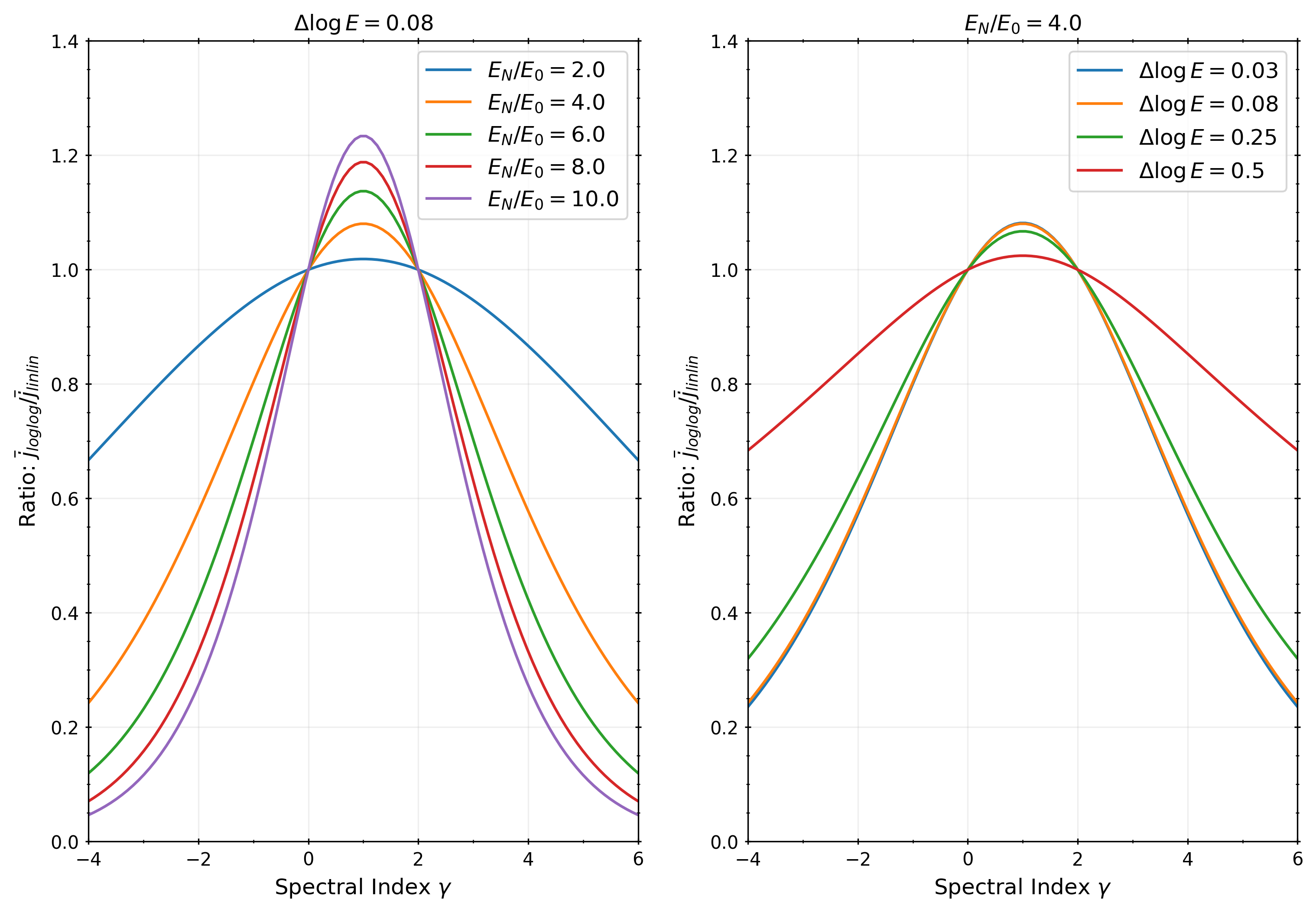}
    \caption{The ratio \(\overline{j}_{\rm loglog}/\overline{j}_{\rm linlin}\) defined in Eq. \eqref{eq:ratio} as a function of spectral index. The left panel plots the ratio while varying \(E_N/E_0\in \{2,4,6,8,10\}\) for a constant \(\Delta \log E = 0.08\), whereas the right panel plots the ratio for a constant \(E_N/E_0=4\) while varying \(\Delta \log E \in \{0.03,0.08,0.25,0.5\}\).}
    \label{fig:modeled_ratio}
\end{figure}

In Figure \ref{fig:modeled_ratio}, we show values of Eq. \eqref{eq:ratio} for combinations of \(E_N / E_0\) and \(\Delta \log E\) for intensity spectra following the power law \(E^{-\gamma}\) with spectral indices ranging from -4 to 6 (\(\gamma<0\) for rising spectra, \(\gamma=0\) for flat spectra, and \(\gamma>0\) for falling spectra).
SEP intensities as a function of energy are often characterized by falling spectra, with spectral indices observed by PSP ranging between 0 and 7 \citep{MitchellEA2023ApJS_ISOIS_LivingCatalog}.
Rising spectra are included in the modeled results of Figure \ref{fig:modeled_ratio} for symmetry; although, their results are applicable for special circumstances \citep[e.g., during a transition period from SEPs to cosmic ray intensities]{GiacaloneEA2022SSRv_CosmicRays_transport_energization}.  
% (increasing spectra (\(\gamma < 0\)) are characteristic of cosmic rays \citep[e.g., ][]{RankinEA2021ApJ_ACRS_PSP} so they are examined here as well).
These spectra are exact, absent of random sampling errors, and the energy bins are constructed to have equal width in logarithmic space.
We see that only for the special cases of \(\gamma=0\) and \(\gamma=2\) do we have that \(\overline{j}_{\rm loglog}/\overline{j}_{\rm linlin} = 1\).
They are also approximately equal when the new energy range is small but are very different with a large energy range (left panel of Figure \ref{fig:modeled_ratio}).
With \(\Delta \log E=0.08\) fixed (small original energy bins) while varying \(E_N/E_0\in \{2,4,6,8,10\}\) (increasing new energy range), the value of \(\overline{j}_{\rm linlin}\) can quickly become greater than 2 times the value of \(\overline{j}_{\rm loglog}\), as in the case \(E_N/E_0=4\) and \(\gamma=5\).
The right panel of Figure \ref{fig:modeled_ratio} shows the expected ratio as a function of spectral index for a constant \(E_N/E_0=4\) while varying values of \(\Delta \log E \in \{0.03,0.08,0.25,0.5\}\), which confirms that the ratio is largely dependent on the spectral index and \(E_N/E_0\) for a sufficiently small energy bin.

% \newpage

\section{Comparisons with Observations} \label{sec:results}

\subsection{Data Description} \label{subsec:data}

For the results in Section \ref{subsec:observations}, we apply Eqs. \eqref{eq:j_linlin} and \eqref{eq:j_loglog} to compute \(\overline{j}_{\rm linlin}\) and \(\overline{j}_{\rm loglog}\), respectively, to energetic proton intensity-energy spectra measured by the PSP/IS\(\odot\)IS instrument suite, composed of EPI-Lo and EPI-Hi \citep{McComasEA2016SSRv_ISOIS,HillEA2017JGRA_EPILo,WiedenbeckEA2021AandA_ISOIS_EPIHI}.
The data is publicly available at NASA's Space Physics Data Facility\footnote{Space Physics Data Facility can be accessed at \url{https://cdaweb.gsfc.nasa.gov/}.}.%{https://cdaweb.gsfc.nasa.gov/}.}.
EPI-Hi has a total of five different broad look-directions, three covered by the low energy telescopes (LET) and two covered by the high energy telescope (HET).
LET-A/HET-A and LET-B/HET-B are the ``sun-ward" and ``anti sun-ward" facing aperture pairs, respectively, where LET-C looks perpendicular to LET-A and LET-B.
EPI-Lo is composed of eight wedges for a total of 80 different narrower look-directions, with a combined field-of-view covering nearly half of the sky.
In the present analysis, we take an omni-directional average of the ChanP proton intensity observed by EPI-Lo, excluding apertures that have been impacted by dust and thus have high noise backgrounds \citep{SzalayEA2020ApJS_PSP_DustImpact,ShenEA2024_dust}.
The total number of energy bins and the range of energies covered by EPI-Lo, LET, and HET (determined from public data) are as follows: 38 bins between 67 -- 8,736~keV for EPI-Lo ChanP protons with \(\Delta \log E\) ranging from \(\sim\)~0.03 -- 0.07, 25 bins between 0.6 -- 45.3~MeV for LET protons with \(\Delta \log E\sim0.075\), and 15 bins between 6.7 -- 90.5~MeV for HET protons with \(\Delta \log E\sim0.075\). 
In Appendix \ref{app:data}, we describe the methods used for processing IS\(\odot\)IS data concerning the averaging of EPI-Lo look-directions (technique described by Figure \ref{fig:interpolation}) and the Poisson upper and lower limits used for errors \citep{Gehrels1986ApJ_PoissonStatistics}.

\subsection{Observations} \label{subsec:observations}

To examine the difference between \(\overline{j}_{\rm linlin}\) and \(\overline{j}_{\rm loglog}\) (Eqs. \eqref{eq:j_linlin} and \eqref{eq:j_loglog}) in observations, we select a period consisting of a series of SEP events observed by PSP, which includes three-stage particle acceleration of SEPs by a coronal mass ejection driven shock \citep{ChenLi2024ApJL_SEPs_PSP_ThreeStageAcceleration}.
% my be characterized as an enhancement of energetic protons with energies less than \(\sim 20\)~MeV.
These SEP events occur on 26 -- 30 August 2022, which has also been identified as a period with increased \({\rm ^3He}\) abundance observed by Solar Orbiter \citep{KouloumvakosEA2023ApJ_3He_rich_events_SolO}; although, PSP was magnetically connected to the other side of the Sun \citep{GieselerEA2023FrASS_SolarMach}.
Nonetheless, the Sun was undergoing numerous processes at different solar longitudes, making this period particularly interesting.

\begin{figure}[ht]
    \centering
    \includegraphics[width=0.65\linewidth]{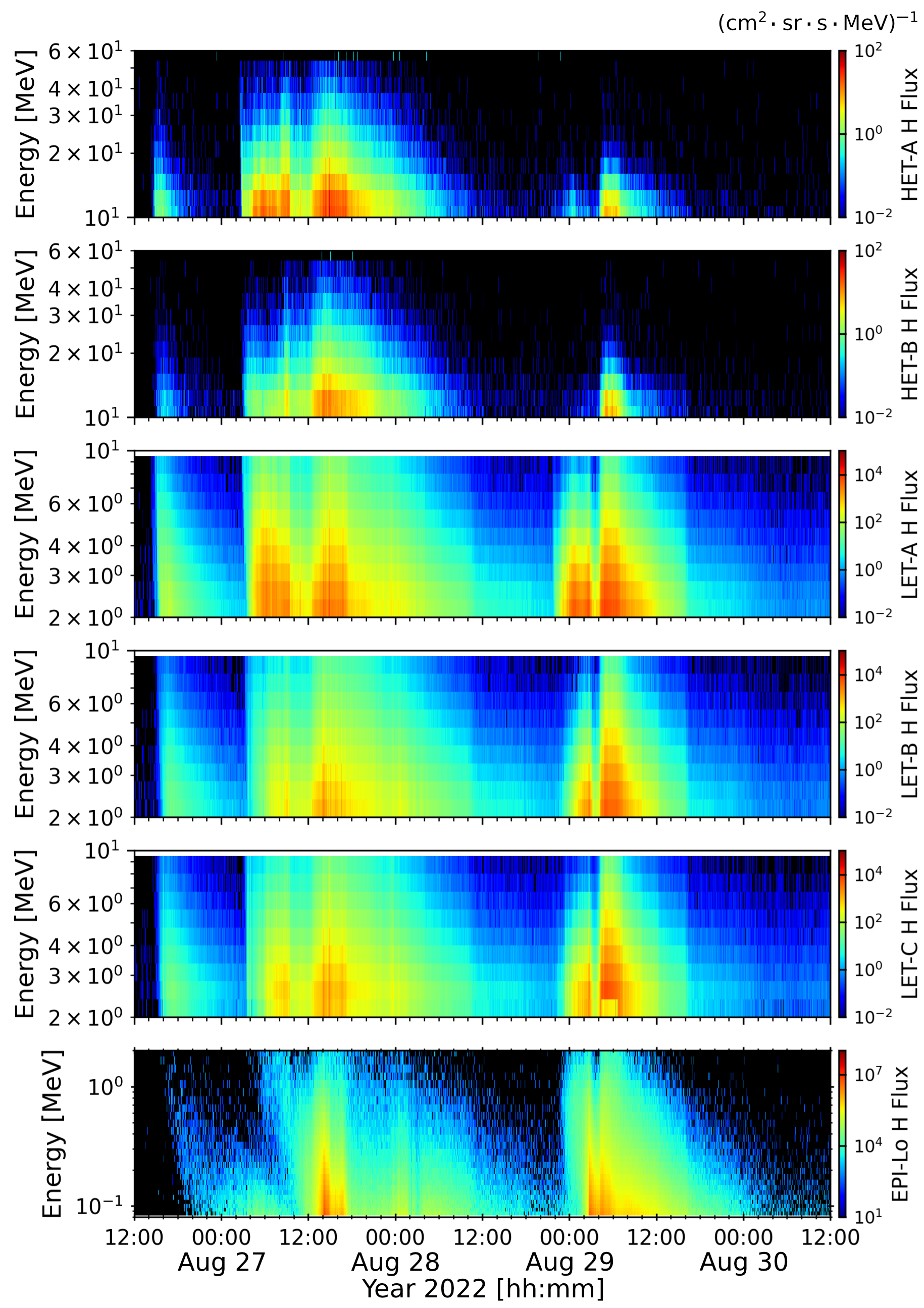}
    \caption{The spectrogram of particle intensity for HET-A, HET-B, LET-A, LET-B, LET-C, and EPI-Lo in descending order at a 1-minute resolution.}
    \label{fig:overview}
\end{figure}

\begin{figure}[ht]
    \centering
    \includegraphics[width=0.5\linewidth]{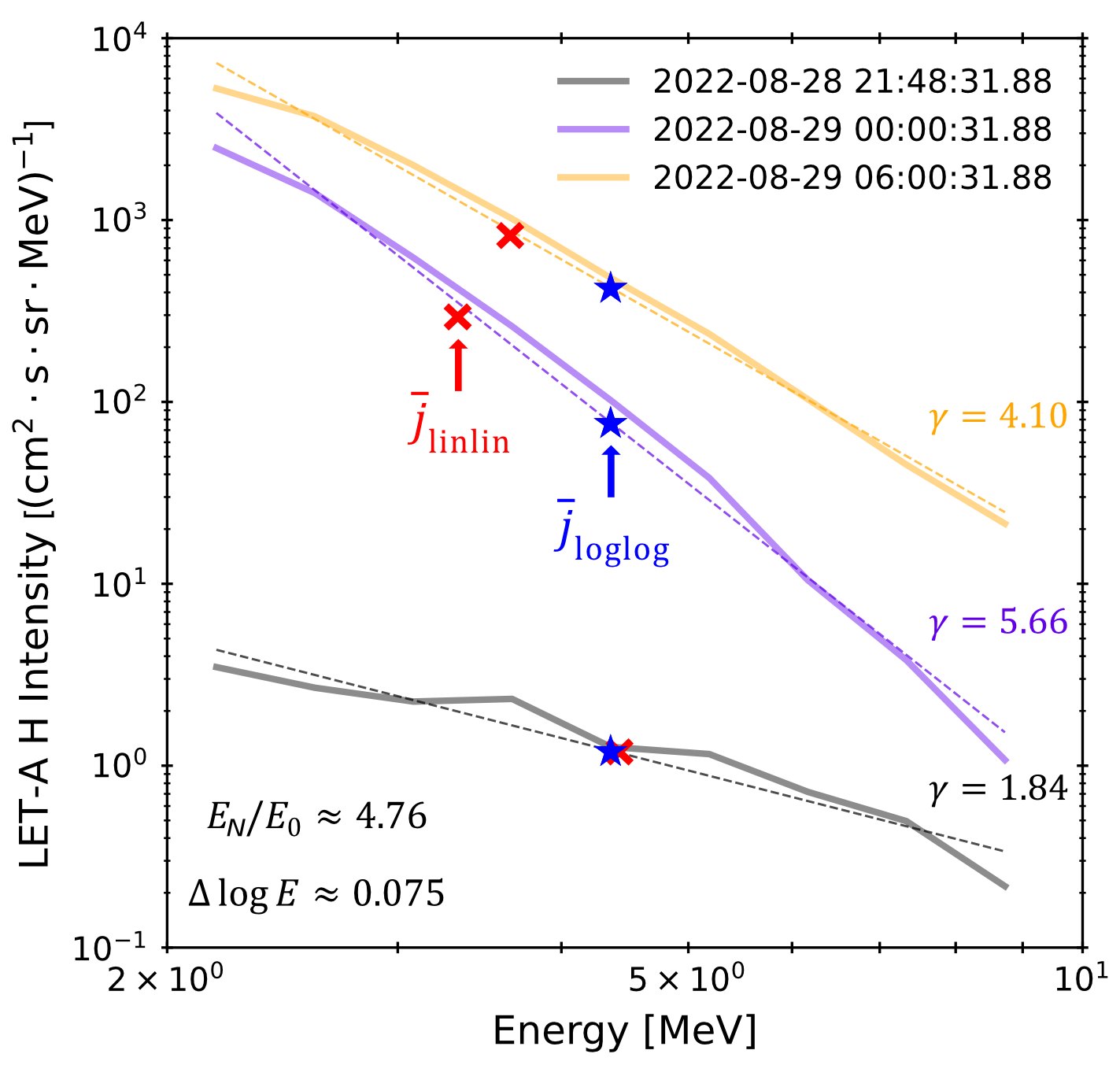}
    \caption{Proton spectra observed by LET-A from 2 -- 10~MeV for three different 1-minute samples during the 28 August 2022 SEP event. The red-\(\times\) and blue-\(\star\) mark the values of \(\overline{j}_{\rm linlin}\) and \(\overline{j}_{\rm loglog}\), respectively. The dashed lines represent linear fits of the spectra in logarithmic space (uncertainties are ignored), with fitted values of the spectral index provided accordingly.}
    \label{fig:sample_spectra}
\end{figure}

In Figure \ref{fig:overview}, we show the spectrogram of proton intensities at 1-minute resolution for each instrument composing the IS\(\odot\)IS suite (EPI-Lo, LET-A, LET-B, LET-C, HET-A, and HET-B), providing an overview of period consisting of a series of SEP events.
At three separate times, we examine the spectrum observed by LET-A in Figure \ref{fig:sample_spectra}, along with their corresponding values of \(\overline{j}_{\rm linlin}\) and \(\overline{j}_{\rm loglog}\).
The dashed lines represent the linear fit of the spectra in logarithmic space (errors in the intensity were ignored for this demonstration), with fitted values of the spectral index provided accordingly.
We can see that in all three cases, the value of \(\overline{j}_{\rm loglog}\) picks out the intensity of the fit at the log-centered energy.
This means that in the presence of nonlinear behavior of the spectrum in logarithmic space, \(\overline{j}_{\rm loglog}\) represents the intensity of a single power law fit at the log-centered energy.
Furthermore, the values of \(\overline{j}_{\rm linlin}\) almost describe the intensities along the best fit at the effective energy, which is found to vary by \(\sim 1\)~MeV for these spectral indices.

\newpage

\newpage

\newpage
\begin{figure}[htp!]
    \centering
    \includegraphics[width=\linewidth]{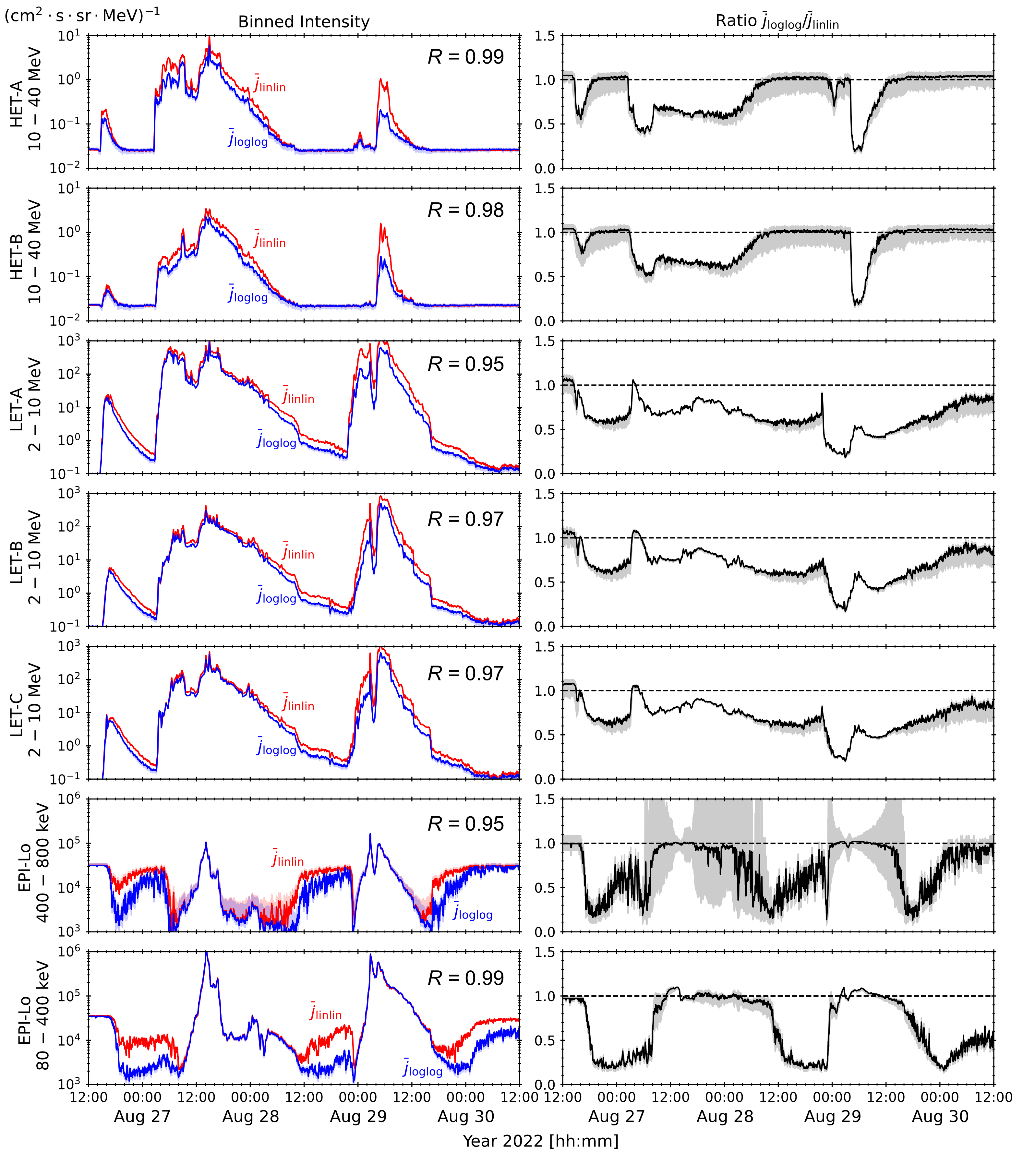}
    \caption{The values of \(\overline{j}_{\rm linlin}\) (red) and \(\overline{j}_{\rm loglog}\) (blue) on the left column for HET-A, HET-B, LET-A, LET-B, LET-C, and EPI-Lo in descending order, smoothed to 11 minutes to remove sharp variations over shorter time scales. On the right column is the ratio \(\overline{j}_{\rm loglog}/\overline{j}_{\rm linlin}\) for the corresponding instrument/energy range. The shaded regions about each curve represent the range of error. The correlation coefficient (\(R\)) between \(\overline{j}_{\rm linlin}\) and \(\overline{j}_{\rm loglog}\) are given in each panel of the left column accordingly.}
    \label{fig:results}
\end{figure}

For each of the spectrograms shown in Figure \ref{fig:overview}, we apply Eqs. \eqref{eq:j_linlin} and \eqref{eq:j_loglog} to each measured spectrum at each time to compute values of \(\overline{j}_{\rm linlin}\) and \(\overline{j}_{\rm loglog}\), respectively, for the following energy ranges: 100~--~400~keV and 400~--~800~keV for EPI-Lo, 2~--~10~MeV for LET, and 10~--~40~MeV for HET.
These results are given in Figure \ref{fig:results} (left column), in addition to their ratios (right column).
Note that the errors of \(\overline{j}_{\rm linlin}\) and \(\overline{j}_{\rm loglog}\) (shaded regions about each curve) are larger in regions characteristic of low to zero count measurements, which are also reflected with larger uncertainty in their ratio.
Upper and lower uncertainties are propagated individually using the inverse variance method.
We report that the time series of \(\overline{j}_{\rm linlin}\) and \(\overline{j}_{\rm loglog}\) are strongly correlated, with correlation coefficients (\(R\)) given in each panel of the left column of Figure \ref{fig:results}.
This means that the general evolution of the proton intensity of a particular SEP event is captured when using either binning procedure.
However, their correlation generally points to a similar tendency in their trends but does not measure the difference in their magnitudes, which can vary greatly as shown by their ratio provided in the right column of Figure \ref{fig:results}.
It is evident from Figure \ref{fig:results} that the difference between \(\overline{j}_{\rm linlin}\) and \(\overline{j}_{\rm loglog}\)is both time and energy dependent, as expected from the dependence of the effective energy on the spectral index found analytically in Section \ref{subsec:analytic}.
For these SEP events, using \(\overline{j}_{\rm linlin}\) to represent the proton intensity at the log-centered energy would consistently yield a larger intensity, and at some times by nearly a factor of five, given any energy range investigated here.

\newpage
\section{Discussion \& Conclusion} \label{sec:discussion}

The average particle intensity is an important quantity used to characterize particles over a specified energy range.
Conventionally, an energy averaged intensity is reported, corresponding to the re-binned intensity (\(\overline{j}_{\rm linlin}\); Eq. \ref{eq:j_linlin}).
This represents a statistical re-binning procedure by returning to count space, summing up the counts over the merged energy bins, and then dividing by the energy width of the newly constructed, wider energy bin.
However, the re-binned intensity does not correspond to the log-centered energy of the new energy bin.
Instead, it corresponds to the value of the spectrum at the effective energy \citep{KronbergDaly2013GIMDS_WideEnergySpectralAnalysis}, which is found to vary in time due to its dependence on the spectral index.
Nonetheless, the log-centered energy is often used to represent \(\overline{j}_{\rm linlin}\) because of its convenience when performing fits in logarithmic space.
% , where the intensity-energy spectrum is expected to mostly behave linearly (excluding transition regions of exponential roll-over or broken power law behavior).
In this paper, we demonstrate that an alternative measure is available for particle observations, which can be used to calculate the intensity with respect to the central energy of the range in logarithmic space, namely the spectral binned intensity.
In other words, this spectral binned intensity, defined as \(\overline{j}_{\rm loglog}\) (Eq. \eqref{eq:j_loglog}), is the intensity along the originally measured power law spectrum at the log-centered energy of the new energy bin resulting from the combination of the originally constructed energy bins.
This allows for spectral comparisons at the same energy, a property that is lost with \(\overline{j}_{\rm linlin}\).

\begin{figure}[ht]
    \centering
    \includegraphics[width=0.85\linewidth]{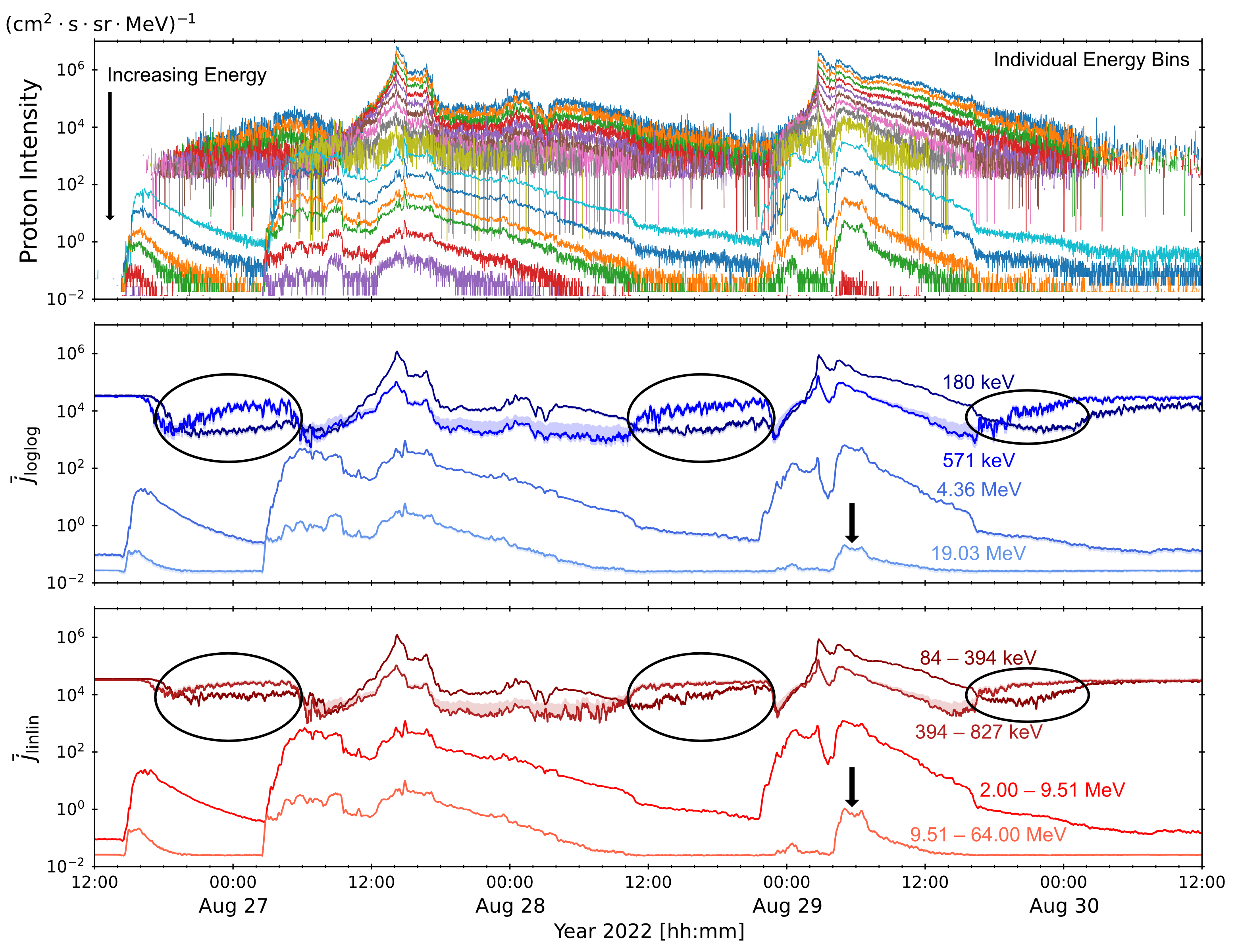}
    \caption{The time series of proton intensities for individual energy bins measured by EPI-Lo, LET-A, and HET-A (top panel), in addition to their corresponding spectral binned intensities \(\overline{j}_{\rm loglog}\) (middle panel) and re-binned intensities \(\overline{j}_{\rm linlin}\) (bottom panel).}
    \label{fig:time_series_comparison}
\end{figure}

Figure \ref{fig:time_series_comparison} shows a comparison of the results from Figure \ref{fig:results} in a different way as a time series.
The top panel consists of every third individual energy bin that is observed from EPI-Lo, LET-A, and HET-A, which is made to be difficult to understand on purpose.
This is a situation where merging energy bins to reduce the number of curves for a visually appealing and scientifically meaningful plot becomes useful.
Therefore, we combine bins over energies as shown in Figure \ref{fig:results} and calculate \(\overline{j}_{\rm loglog}\) and \(\overline{j}_{\rm linlin}\) shown in the middle and bottom panels of Figure \ref{fig:time_series_comparison}, respectively.
At first glance, the general evolution of the intensity is captured, but there are different behaviors between \(\overline{j}_{\rm loglog}\) and \(\overline{j}_{\rm linlin}\) for the wider EPI-Lo energy bins (as indicated by three circled regions in the middle and bottom panels).
The first and second circled areas show that the difference in the intensity between the two new energy bins are larger when using \(\overline{j}_{\rm loglog}\) compared to \(\overline{j}_{\rm linlin}\).
% \textbf{
Analyzing the difference in the intensity between the lower and higher energy bins provides insight into the originally measured spectrum at a given sample in time.
This property is lost when using \(\overline{j}_{\rm linlin}\) as opposed to \(\overline{j}_{\rm loglog}\).
% }
The third circled area points out that the intensity at higher energy overcomes the intensity at lower energy earlier (\(\sim 1.5~{\rm hours}\)) when using \(\overline{j}_{\rm linlin}\) compared to \(\overline{j}_{\rm loglog}\).
% \textbf{
In this case, the choice of \(\overline{j}_{\rm linlin}\) or \(\overline{j}_{\rm loglog}\) can significantly impact the time at which the spectrum transitions from a falling to rising spectrum.
This detail is important, for instance, in determining the time of a spectral roll up of energetic storm particles, or when particles with lower energy become more abundant than those with higher energy during a time of velocity dispersion \citep{GiacaloneEA2023ApJ_ESPs_2022Feb16event_rollups}.
% }
Furthermore, we point out that the intensity of the highest energy is noticeably different at the period marked by the black arrow, such that \(\overline{j}_{\rm linlin}\) is larger than \(\overline{j}_{\rm loglog}\).
% \textbf{
This signifies that the abundance of higher energy particles may be reported as having a small or large  abundance with respect to some reference intensity, a result that is affected by the choice of using \(\overline{j}_{\rm linlin}\) or \(\overline{j}_{\rm loglog}\).
% }
Note that when labeling each of these curves in the middle and bottom panels of Figure \ref{fig:time_series_comparison}, values of \(\overline{j}_{\rm linlin}\) can only be reported using a range of energy whereas a specific energy can be reported for values given by \(\overline{j}_{\rm loglog}\).

\newpage

\begin{figure}[ht]
    \centering
    \includegraphics[width=0.55\linewidth]{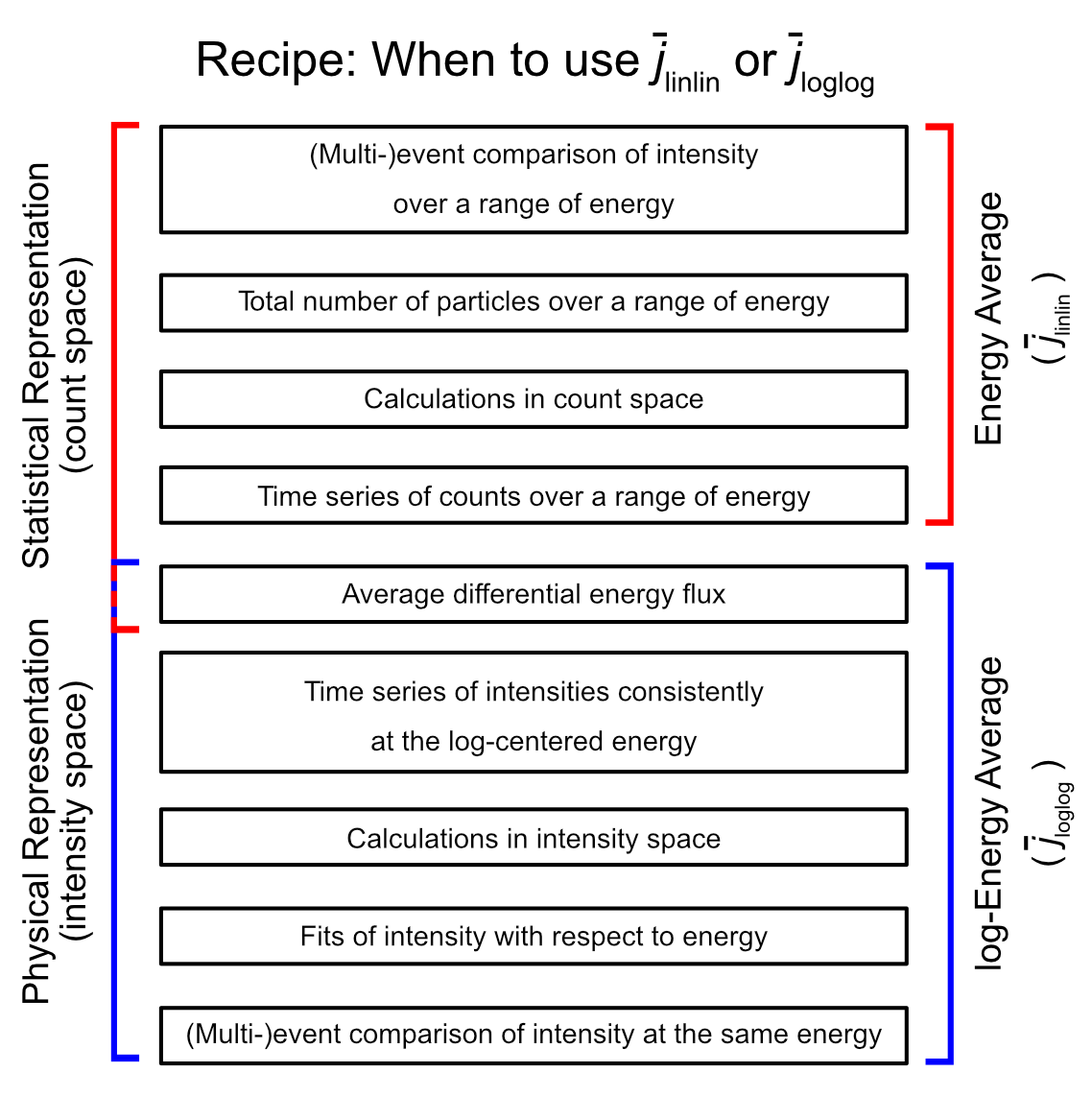}
    \caption{A flowchart on which average measure to use depending on the physical or statistical description desired.}
    \label{fig:recipe}
\end{figure}

Although this property of \(\overline{j}_{\rm loglog}\) is convenient for accurately depicting the intensity at the log-centered energy, there are situations where \(\overline{j}_{\rm linlin}\) is still a more appropriate measure.
Figure \ref{fig:recipe} illustrates different scenarios where averaging in the linear or logarithmic space of energy is more appropriate, which depends on whether a calculation using counts (statistical representation) or intensities (physical representation) is required.
Generally, if it is most physically meaningful to use a consistent energy to characterize the spectrum, then \(\overline{j}_{\rm loglog}\) should be used.
To characterize the spectrum by the number of counts in a wider energy bin, then \(\overline{j}_{\rm linlin}\) should be used.
However, let us consider the case regarding the energy average of the differential energy flux \(\left( \frac{E\cdot N}{G_\textrm{eff}\Delta t\Delta E} = \frac{N}{G_\textrm{eff}\Delta t\Delta \log E} \sim E \cdot j \right)\).
Then, according to Eq. \eqref{eq:average}, \({\rm \hat{X}}(E)=\log E\), \({\rm \hat{Y}}(E\cdot j)=E\cdot j\), and \({\rm xy}={\rm ``loglin"}\) should be used to derive a statistical representation of the average, or calculating \(\overline{E\cdot j}_{\rm loglin}\).
Therefore, an energy average of \(E\cdot j\) in the linear space of \(E\cdot j\) with energy widths described in linear space (\(\overline{E\cdot j}_{\rm linlin}\)) no longer represents a count space operation, losing a statistical description of the averaged differential energy flux.
On the contrary, \(\overline{E\cdot j}_{\rm loglog}\) still retains a physical representation of the quantity, representing the value of \(E\cdot j\) at the log-centered energy of the wider energy bin.
As a result, when \({\rm xy}={\rm loglog}\), the physical meaning of the resulting average is invariant with respect to the variable being averaged.

\newpage

In this paper, we demonstrated the importance of using \(\overline{j}_{\rm loglog}\) for comparing intensities at the same log-centered energy in favor of \(\overline{j}_{\rm linlin}\).
However, \(\overline{j}_{\rm linlin}\) is still a useful measure of representing the number of counts over the merged energy bins.
We found an analytical expression for describing the behaviors of \(\overline{j}_{\rm linlin}\), \(\overline{j}_{\rm loglog}\), and their ratio, as well as expressions for the energies at which they characterize the original intensity spectrum for the case of a single power law distribution.
The same analysis can be used to examine more complex distributions, such as exponential or multiple power laws.
The effective energy for describing \(\overline{j}_{\rm linlin}\) is dependent on the spectral index, and as a result, changes over time.
On the contrary, \(\overline{j}_{\rm loglog}\) describes the originally observed spectrum at the log-centered energy, which is time invariant.
Assuming that these two quantities are used to characterize the log-centered energy of the merged energy bins (as is common practice), we show that for practical values of \(\Delta \log E\), energy range of the merged bins, and spectral index, that \(\overline{j}_{\rm linlin}\) is consistently larger than the \(\overline{j}_{\rm loglog}\) by nearly a factor of five, as demonstrated with PSP/IS\(\odot\)IS measurements.
In this practical demonstration, we find that the time evolution of the intensity is qualitatively conserved given their strong correlation.
The application of \(\overline{j}_{\rm linlin}\) and \(\overline{j}_{\rm loglog}\) is not constrained to energetic particle populations, as was the focus of this paper regarding observations.
However, the intensity-energy spectrum of the bulk solar wind deviates from power-law behavior, unlike the case for SEPs that are often described by power laws.
Thus, further work includes improving the procedure of logarithmic averaging in the presence of zero count/intensity samples and its application to other types of distributions as mentioned above.
This work clarifies the necessity of carefully choosing the appropriate measure of particle intensity regarding the corresponding scientific question and the underlying meaning of the measure itself.

%------------------------------------------------------
\section{Acknowledgments}

We thank the IS\(\odot\)IS team and everyone that made the PSP mission possible. The IS\(\odot\)IS data and visualization tools are available to the community at \href{https://spacephysics.princeton.edu/missions-instruments/PSP}{https://spacephysics.princeton.edu/missions-instruments/PSP}. 
PSP was designed, built, and is operated by the Johns
Hopkins Applied Physics Laboratory as part of NASA’s Living with a Star (LWS) program (contract NNN06AA01C).

\newpage
\appendix
\counterwithin{figure}{section}
\section{Data Processing}\label{app:data}

EPI-Lo was constructed with a total of 80 apertures or look-directions, six of which were constructed with thick foils, by design, to decrease the background from scattered light during the encounter period \citep{McComasEA2016SSRv_ISOIS}, thus resulting in higher initial energy bins compared to the other 74 thinner foil apertures.
These look-directions are identified as 23, 30, 31, 32, 33, and 40.
To account for these shifted energies when combining intensities between look-direction, we interpolate the intensity from the thick foil energies onto the grid of energy bins characteristic to the thin foil apertures using a linear, point-to-point fit in logarithmic space.
We checked other fitting methods, such as spline fitting of orders three and five; however, using these other fitting methods resulted in intensities that largely deviate from the observed spectrum, and hence are undesirable.
The same procedure is applied to the asymmetric errors from the upper and lower limits \citep[][; see their Eqs. (9) and (12) using a 84.13\% confidence level).]{Gehrels1986ApJ_PoissonStatistics}.
As a result, regular linear regression should not be used for weighted fits of these newly interpolated data. Instead, other regression models, such as Poisson regression \citep{Nelder1974JRSS_PoissonRegression} should be used.

\begin{figure}[ht]
    \centering
    \includegraphics[width=0.4\linewidth]{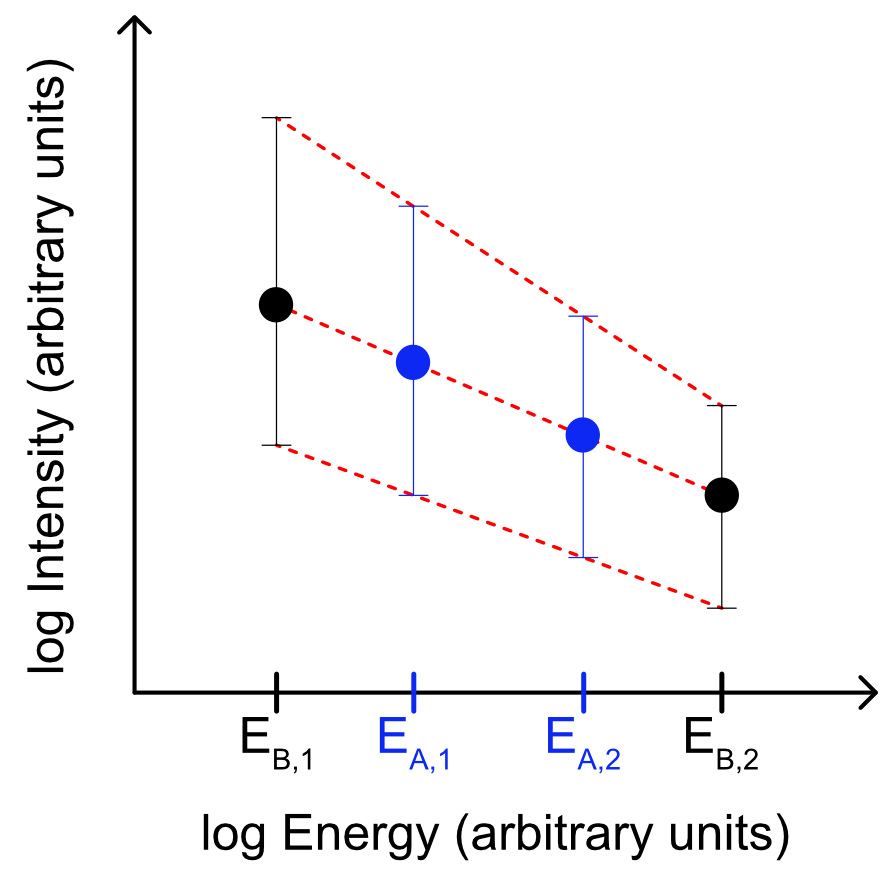}
    \caption{An illustration demonstrating the procedure for interpolation of intensity from the energy grid of thick to thin foil EPI-Lo apertures. Thin and thick foil energies are sub-scripted with ``A" and ``B", respectively. The black points and their error bars represent the intensity and its asymmetric error for two thick foil energy bins. The blue points and their error bars represent the interpolated intensity and error from the thick foil intensities onto the thin foil energy grid that may exist anywhere between the adjacent thick foil energies.}
    \label{fig:interpolation}
\end{figure}

In Figure \ref{fig:interpolation}, we show an illustration of this procedure to interpolate the intensities from thick foil apertures onto the energy grid of thin foil apertures. 
To handle zeros before interpolation in logarithmic space, we first add 1 intensity unit to all measured intensities, perform the interpolation in logarithmic space, and finally subtract 1 intensity unit from the newly interpolated values.
The steps of adding/subtracting 1 intensity unit does not need to be applied to the errors since, in logarithmic space, a zero error is still zero, and this operation does not alter the originally measured uncertainty.
This conveniently allows us to describe the behavior of the spectrum in the presence of zeros while avoiding the logarithm of zero.
We investigated other values to modify the values of the intensity before taking its logarithm; however, the change in the distribution of resulting intensities after interpolation was insignificant.

\newpage
    
\bibliographystyle{plainnat}
% \begin{thebibliography}{}
% \include{CuestaBibItems}
% \end{thebibliography}
% \bibliography{CuestaBib}

%% This command is needed to show the entire author+affilation list when
%% the collaboration and author truncation commands are used.  It has to
%% go at the end of the manuscript.
%\allauthors

%% Include this line if you are using the \added, \replaced, \deleted
%% commands to see a summary list of all changes at the end of the article.

%\listofchanges

\end{document}